\documentclass[a4paper,11pt]{article}
\usepackage{jcappub} 
\usepackage{lineno}
\usepackage{gensymb}
\usepackage{comment}
\usepackage{subfigure}
\usepackage{orcidlink}

\newcommand{\gsim}{\raisebox{-0.13cm}{~\shortstack{$>$ \\[-0.07cm]
      $\sim$}}~}

\title{\boldmath Ultrahigh-Energy Gamma-Ray Sources Need Not Be Hadronic PeVatrons}

\author[1\orcidlink{0000-0002-0194-7576}]{Zachary Curtis-Ginsberg,\note{Corresponding Author}}
\author[\orcidlink{0009-0004-2456-1221}]{Dan Hooper,}
\author[\orcidlink{0000-0002-9867-6548}]{and Justin Vandenbroucke}
\affiliation{Department of Physics and Wisconsin IceCube Particle Astrophysics Center, University of Wisconsin, Madison, WI 53706, USA}

\emailAdd{curtisginsbe@wisc.edu}

\abstract{Ultrahigh-energy gamma rays ($E_{\gamma}>100 \, {\rm TeV}$) have been detected from a handful of astrophysical sources. Due to the Klein-Nishina suppression of inverse Compton scattering at such high energies, it has sometimes been argued that these sources must be accelerators of PeV-scale protons, making them the long-sought-after Galactic ``PeVatrons.'' Here, we challenge this conclusion, demonstrating that these sources can be straightforwardly explained by simple leptonic models. In this context, we consider the microquasar SS 433, the Galactic Center, and TeV halos, showing in each case that the observation of PeV-scale gamma rays from these sources does not indicate that they are accelerators of hadronic cosmic rays. We also note that the measured angular extension of SS 433 is in good agreement with the predictions of our model, favoring a leptonic origin for the gamma-ray emission from this source. A definitive identification of a PeVatron would require additional information, such as the combined observation of the pion bump and synchrotron peak, the spatial correlation of gamma-ray emission with gas, or the detection of neutrinos with $E_{\nu} \gsim 100 \, {\rm TeV}$.}

\begin{document}
\maketitle
\flushbottom

\section{Introduction} \label{intro}

Much regarding the origin of the cosmic rays remains a mystery, although the shape and chemical composition of the cosmic-ray spectrum provide clues about where these particles may have been accelerated~\cite{Hoerandel:2004gv,Allard:2005ha}. These and other considerations suggest that the highest-energy cosmic rays originate from extragalactic sources, while the lower-energy portion of this spectrum is dominated by Galactic sources. Candidates for Galactic cosmic-ray accelerators include supernova remnants~\cite{AGILE:2010egi,Fermi-LAT:2013iui,Jouvin:2017bwo}, microquasars~\cite{Kaci:2025gyb}, pulsar wind nebulae~\cite{Arons:2012tz,Gaensler:2006ua}, star-forming regions~\cite{Bykov:2020zqf}, massive stars~\cite{Aharonian:2018oau}, and the Milky Way's central supermassive black hole~\cite{Fujita:2016yvk,Guo:2016zjl,HESS:2016pst}.

At an energy near $E\sim 4 \, {\rm PeV}$, the spectral index of the cosmic-ray spectrum softens from approximately 2.7 to 3.1. It has long been suggested that this spectral feature, known as the ``knee,'' could be associated with the maximum energy to which Galactic sources are able to accelerate cosmic-ray protons. Measurements further indicate that the composition of the cosmic-ray spectrum becomes heavier at energies between the knee and $\sim 10^2 \, {\rm PeV}$~\cite{Apel:2013uni,EAS-TOP:2004aim,KASCADE:2005ynk,IceCube:2012vv,LHAASO:2024knt}, consistent with a scenario in which the maximum energy of Galactic cosmic-ray acceleration is proportional to the charge of the nuclear species, $E^{\rm max}_Z \approx Z\,E_p^{\rm max}$\cite{Hillas:2005cs,Stanev:1993tx,Kobayakawa:2000nq} (for a review, see Ref.~\cite{Kachelriess:2019oqu}).

Some of the strongest evidence for hadronic cosmic-ray acceleration in the Milky Way comes from multi-wavelength studies of the supernova remnants IC 443 and W44~\cite{AGILE:2010egi,Fermi-LAT:2013iui}, which indicate that these sources accelerate protons to TeV-scale energies. Young supernova remnants, such as Cas A~\cite{MAGIC:2017hoy} and Tycho's supernova remnant~\cite{VERITAS:2017qhy}, are also well fit by hadronic models and appear likely to accelerate protons to somewhat higher energies. They do not, however, seem to accelerate particles to the PeV scale. This conclusion is further supported by theoretical considerations which suggest that supernova remnants cannot efficiently accelerate protons to the energy of the knee~\cite{Lagage:1983zz,Cristofari:2020mdf,Gabici:2019jvz,Gabici:2016fup,Bell:2013kq}. 

In recent years, observations of ultrahigh-energy (UHE) gamma rays from several Galactic sources have been cited as evidence that these sources are accelerators of PeV-scale protons, making them Galactic ``PeVatrons.'' The H.E.S.S. Collaboration presented such an argument in the context of the Milky Way's central supermassive black hole, Sgr A*~\cite{HESS:2016pst}, and analogous statements were made by the HAWC Collaboration after reporting $\gsim 100 \, {\rm TeV}$ gamma-ray emission from three TeV halos~\cite{HAWC:2019tcx}. More recently, the LHAASO Collaboration has made similar claims regarding the microquasar SS 433~\cite{LHAASO:2024psv} and the star-forming region Cygnus X~\cite{LHAASO:2023uhj,LHAASO:2025ysm}. We argue that these observations of UHE gamma rays are not sufficient to claim that a source is a PeVatron.

The remainder of this paper is structured as follows. Sec.~\ref{leptonic_model} describes the calculation of the inverse Compton spectrum from a generic leptonic source. In Sec.~\ref{pevatron_examples}, we describe three examples of possible PeVatron candidates and demonstrate that their observed features can each be accommodated by leptonic processes, with particular attention given to the microquasar SS 433. Sec.~\ref{shapes} outlines generic examples to compare observations with. Finally, in Sec.~\ref{conclusion}, we summarize our results and conclusions.

\section{Ultrahigh-Energy Gamma Rays from Inverse Compton Scattering} \label{leptonic_model}

In this section, we describe our calculations of the gamma-ray emission produced through the leptonic process of inverse Compton scattering. The propagation and energy losses of high-energy electrons are modeled according to the following transport equation:
\begin{align} \label{diff_eq}
    \frac{\partial}{\partial t}\frac{dn_e}{dE_e}(E_e, r, t) =&
    \vec\nabla \cdot \left[ D(E_e) \vec\nabla \frac{dn_e}{dE_e}(E_e, r, t)  \right] 
    + \frac{\partial}{\partial E_e} \left[ \frac{dE_e}{dt}(r) \frac{dn_e}{dE_e}(E_e, r, t) \right] + \delta (r) Q(E_e,t),
\end{align}
where $dn_e/dE_e$ is the differential number density of electrons, $r$ is the distance from the source, $D$ is the diffusion coefficient, and $Q$ defines the spectrum and time profile of the electrons injected from the source. The electrons undergo both inverse Compton scattering and synchrotron radiation, leading to the following energy loss rate~\cite{Blumenthal:1970gc}:
\begin{align}
    \frac{dE_e}{dt} \approx - \sum_i \frac{4}{3} \sigma_t \rho_i S_i(E_e) \frac{E^2_e}{m^2_e}  -\frac{4}{3}\sigma_t \rho_{\rm mag} \frac{E^2_e}{m^2_e},
\end{align}
where $\sigma_t$ is the Thompson scattering cross section and $\rho_{\rm mag}= B^2/2\mu_0$ is the energy density of the magnetic field. The sum in this expression is performed over the various thermal components of the radiation background, and $\rho_i$ and $S_i$ are the energy density and Klein-Nishina suppression factor associated with each such component. The Klein-Nishina suppression factor can be approximated by~\cite{Blumenthal:1970gc}
\begin{align}
S_i(E_e) \approx \frac{45 m^2_e/64\pi^2 T^2_i}{(45 m^2_e/64\pi^2 T^2_i)+(E^2_e/m^2_e)},
\end{align}
where $T_i$ is the temperature of the $i$th radiation component. We adopt a radiation model that consists of the cosmic microwave background ($\rho_{\rm CMB} = 0.260 \, \mathrm{eV/cm}^3$, $T_{\rm CMB} = 2.7\, \mathrm{K}$), infrared emission ($\rho_{\rm IR} = 0.60 \,\mathrm{eV/cm}^3$, $T_{\rm IR} = 20 \, \mathrm{K}$), starlight ($\rho_{\rm star} = 0.60 \,\mathrm{eV/cm}^3$, $T_{\rm star} = 5,000 \, \mathrm{K}$), and ultraviolet emission ($\rho_{\rm UV} = 0.10 \,\mathrm{eV/cm}^3$, $T_{\rm UV} = 20,000 \,\mathrm{K}$)~\cite{Porter:2005qx}.

For a spectrum of injected electrons of the form 
\begin{align}
\label{eq:burst}
Q(E_e, t) = \delta(t) Q_0 E^{- \alpha} \mathrm{exp}[-E_e/E_c], 
\end{align}
the solution to Equation~\ref{diff_eq} is given by 
\begin{align}
\begin{split}
    \frac{dn_e}{dE_e}(E_e, r, t) =& \frac{Q_0 E_0^{2-\alpha}}{8 \pi^{3/2} E_e^2 L_{\rm dif}^3(E_e, t)} \, \mathrm{exp} \left[ \frac{-E_0}{E_c} \right] \: \mathrm{exp} \left[ \frac{-r^2}{4 L_{\rm dif}^2(E_e, t)} \right],
\end{split}
\end{align}
where $E_0$ is the initial energy of an electron which has energy $E_e$ after a time $t$, which is found by solving 
\begin{align}
\int_{E_0}^{E_e} \frac{dE_e'}{dE_e/dt(E_e')} = t,
\end{align}
and the diffusion length scale is defined as 
\begin{align}
    L_{\rm dif}(E_e, t) \equiv \left[ \int_{E_0}^{E_e} \frac{D(E_e')}{dE_e/dt(E_e')}dE_e' \right]^{1/2}.
\end{align}
We parameterize the diffusion coefficient as $D(E_e) = D_0 E_e^\delta$ and adopt $D_0 = 3.86 \times 10^{28} {\mathrm{cm}}^2/{\rm s}$ and $\delta = 1/3$, as is appropriate for cosmic-ray transport throughout the bulk of the Milky Way's interstellar medium (ISM)~\cite{Porter:2021tlr}. To obtain the distribution of electrons for an arbitrary time profile of injection, we perform a sum over the solutions obtained for burst-like time steps (as characterized in Eq.~\ref{eq:burst}). We adopt time steps that are much smaller than the synchrotron cooling time in the relevant energy regime. 

Once we have the spectrum and spatial distribution of the high-energy electrons surrounding a source, we calculate the spectrum and spatial distribution of the gamma-ray emission, $dn_\gamma/dE_\gamma(E_\gamma, r, t)$, using the publicly available Python package, {\it Naima}~\cite{Zabalza:2015bsa,Khangulyan:2013hwa}. We then calculate the observed spectrum and angular distribution of gamma rays by integrating over the line of sight,
\begin{align}
    \frac{dN_\gamma}{dE_\gamma}(E_\gamma, \theta, t) = \int \frac{dn_\gamma}{dE_\gamma}(E_\gamma,l, t)  \, dl,
\end{align}
where the distance along the line of sight, $l$, is related to the distance to the source, $r$, according to the law of cosines, $r^2 = d^2 + l^2 - 2dl \cos{\theta}$, where $d$ is the distance from Earth to the source and $\theta$ is the angle from the center of the source.

\section{PeVatron Candidates} \label{pevatron_examples}

\subsection{Microquasar SS 433} \label{ss433}

SS 433 is a bright X-ray source consisting of a black hole and a main sequence companion star~\cite{Cherepashchuk:2019ibu}, located 4.9 kpc from the Solar System~\cite{Su:2018aug}. It was the first microquasar to be discovered, originally cataloged by Stephenson and Sanduleak in 1977~\cite{Stephenson:1977apr}. SS 433's black hole is undergoing a high rate of accretion and features persistent relativistic jets, distinguishing it from most other microquasars. 

Very-high-energy (VHE) gamma-ray emission has been reported from this source by HAWC ~\cite{HAWC:2024ysp}, H.E.S.S.~\cite{HESS:2024jan}, LHAASO~\cite{LHAASO:2024psv}, and VERITAS~\cite{Klein:2025sep}. While each of these telescopes has observed gamma rays up to tens of TeV associated with the jets of this source, LHAASO has also reported emission in excess of 100 TeV that appears to originate from the central region of this binary system. To investigate the origin of this UHE emission, we compare the spectrum and angular distribution of the gamma-ray emission reported by the LHAASO Collaboration to that predicted by the leptonic model described in Sec.~\ref{leptonic_model}. For a discussion of hadronic models for SS 433, see Refs.~\cite{Carpio:2025,Basanti:2025,Ohira:2024qtr}.

\begin{figure}[t]
\centering
\includegraphics[width=0.9\linewidth]{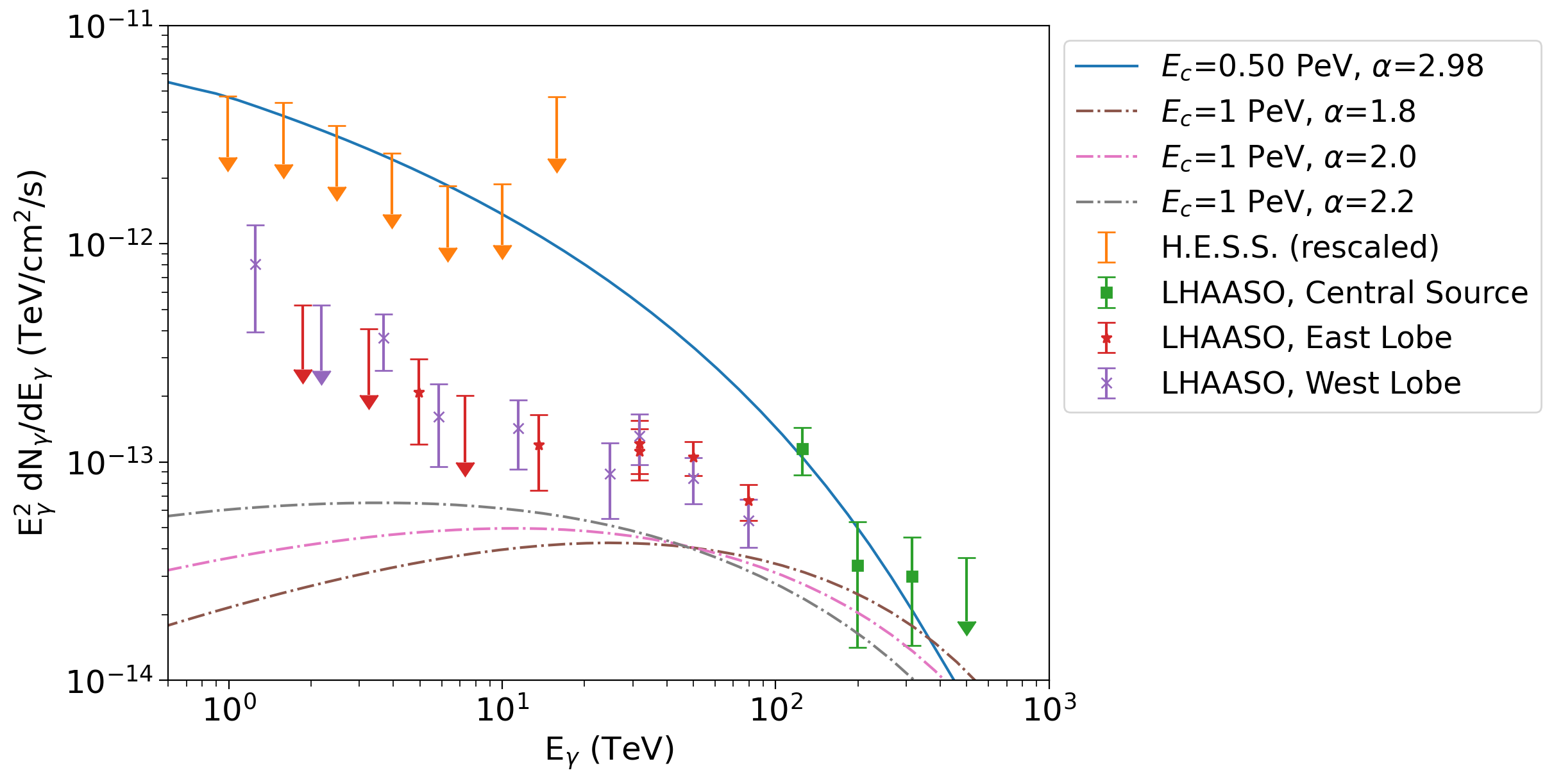}
\caption{The spectrum of the gamma-ray emission from the microquasar SS 433. The flux attributed to the central extended source is shown in green, while the central source upper limits from H.E.S.S. are shown in orange. Based on the spectrum of the central source alone, we obtain the best fit for an injected electron spectral index of $\alpha=2.98$ and a cutoff energy of $E_c=0.50 \, {\rm PeV}$, as shown in solid blue. This model, however, is inconsistent with LHAASO observations at lower energies, which do not reveal any appreciable emission from the central region of this source. For $\alpha \sim 1.8-2.2$ and $E_c \sim 1 {\rm PeV}$, our leptonic model predicts a spectrum (shown as dash-dotted curves) that is broadly consistent with LHAASO's observations, both at relatively low energies and above 150 TeV.  \label{fig:SS433_cent_spec}}
\end{figure}

Since the estimated age of of SS 433 ($\sim$10-100 kyr \cite{Zealey:1980sep,Goodall:2011jul,Han:2020jun}) is much longer than the cooling time for electrons in the energy range of interest, the age and time profile of the electron injection from this source will not appreciably impact our results. Although the strength of the magnetic field at the jet lobes has been inferred from radio and X-ray synchrotron modeling to be of order $\sim 10-20 \,\mu {\rm G}$~\cite{HAWC:2018gwz,Brinkmann:2006zt}, this quantity is largely unknown at the location of the central source, from which LHASSO's UHE emission originates. From energetic considerations, it is clear that such a large magnetic field could not be maintained across the entire region from which LHAASO reports UHE emission.

In Fig.~\ref{fig:SS433_cent_spec}, we compare the predictions of our model to the central source spectrum reported by the LHAASO Collaboration (including the reported statistical and systematic uncertainties), and to the upper limits on the emission from the central source as reported by H.E.S.S. In presenting the gamma-ray spectrum predicted from this source, we show the emission from the entire central source, while the upper limits from H.E.S.S. are on the emission from a region 0.07$\degree$ in radius, centered on the binary. To account for this, we have rescaled the upper limits from H.E.S.S. by a factor equal to the ratio of the total predicted flux to that predicted from the innermost 0.07$\degree$ radius which varies with energy but is $\sim45$. We adopt a magnetic field strength of 3 $\mu$G, consistent with the conditions found in the ISM of the Milky Way. 

Taking the spectrum measured from the central source alone, we obtain the best fit for an injected spectral index of $\alpha=2.98$ and an injected cutoff energy of $E_c=0.50 \, {\rm PeV}$, as shown in solid blue. This model, however, is inconsistent with LHAASO observations at lower energies, which do not reveal any appreciable emission from the central region of this source. In particular, we note in Fig.~1 of Ref.~\cite{LHAASO:2024psv} that the emission observed at $1-25$ TeV and $25-100$ TeV is brighter from the left and right lobes than from the central source. In our Fig.~\ref{fig:SS433_cent_spec}, the blue curve significantly exceeds the emission observed from either lobe, and thus certainly exceeds the upper limits from the central region. To reconcile our model with this data, we adopt a higher value for the cutoff energy, $E_c=1 \, {\rm PeV}$, and a harder spectral index, $\alpha \sim 1.8-2.2$, leading to a spectrum that is consistent with LHAASO's observations, both at relatively low energies and above 200 TeV (although the single flux measurement near 150 TeV cannot be readily accommodated). These models are shown as dash-dotted curves in Fig.~\ref{fig:SS433_cent_spec}.

\begin{figure}[t]
\centering

\includegraphics[width=0.9\linewidth]{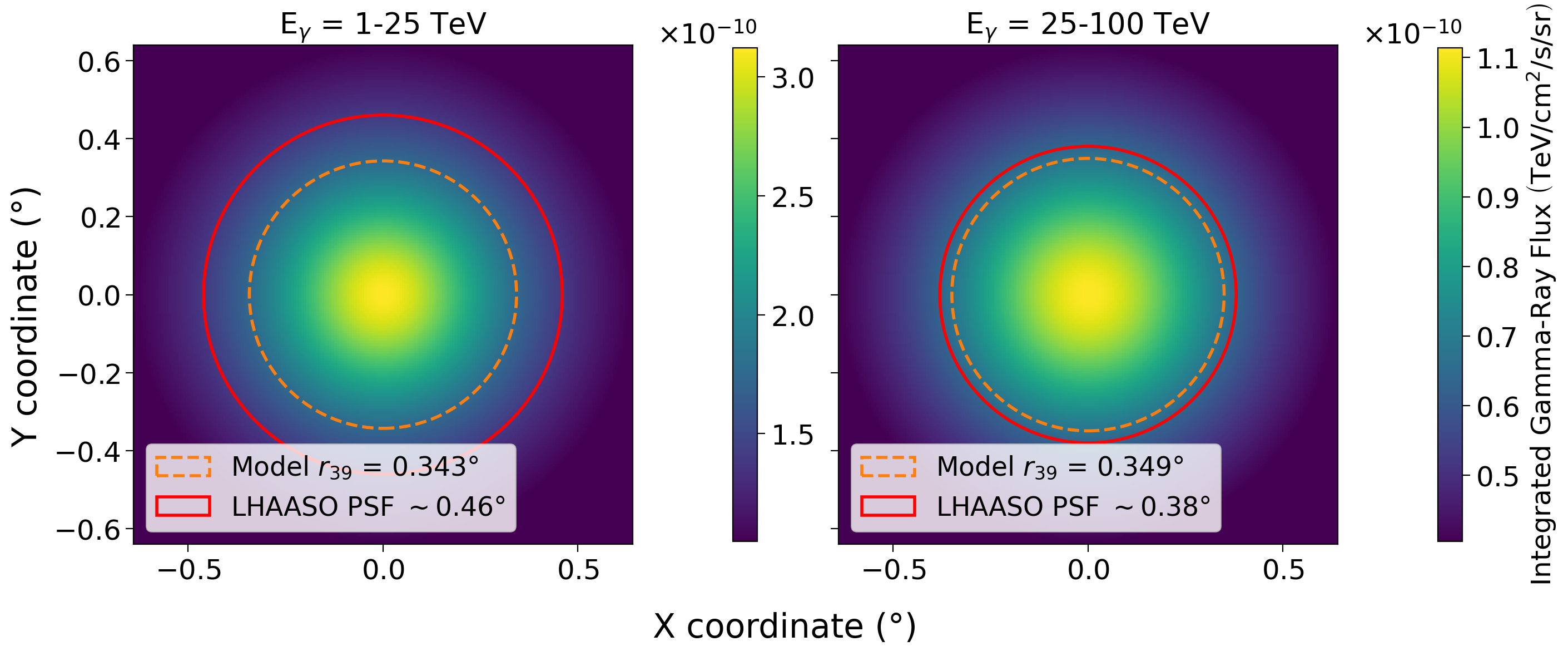}\\
\vspace{0.2cm}
\includegraphics[width=0.5\linewidth]{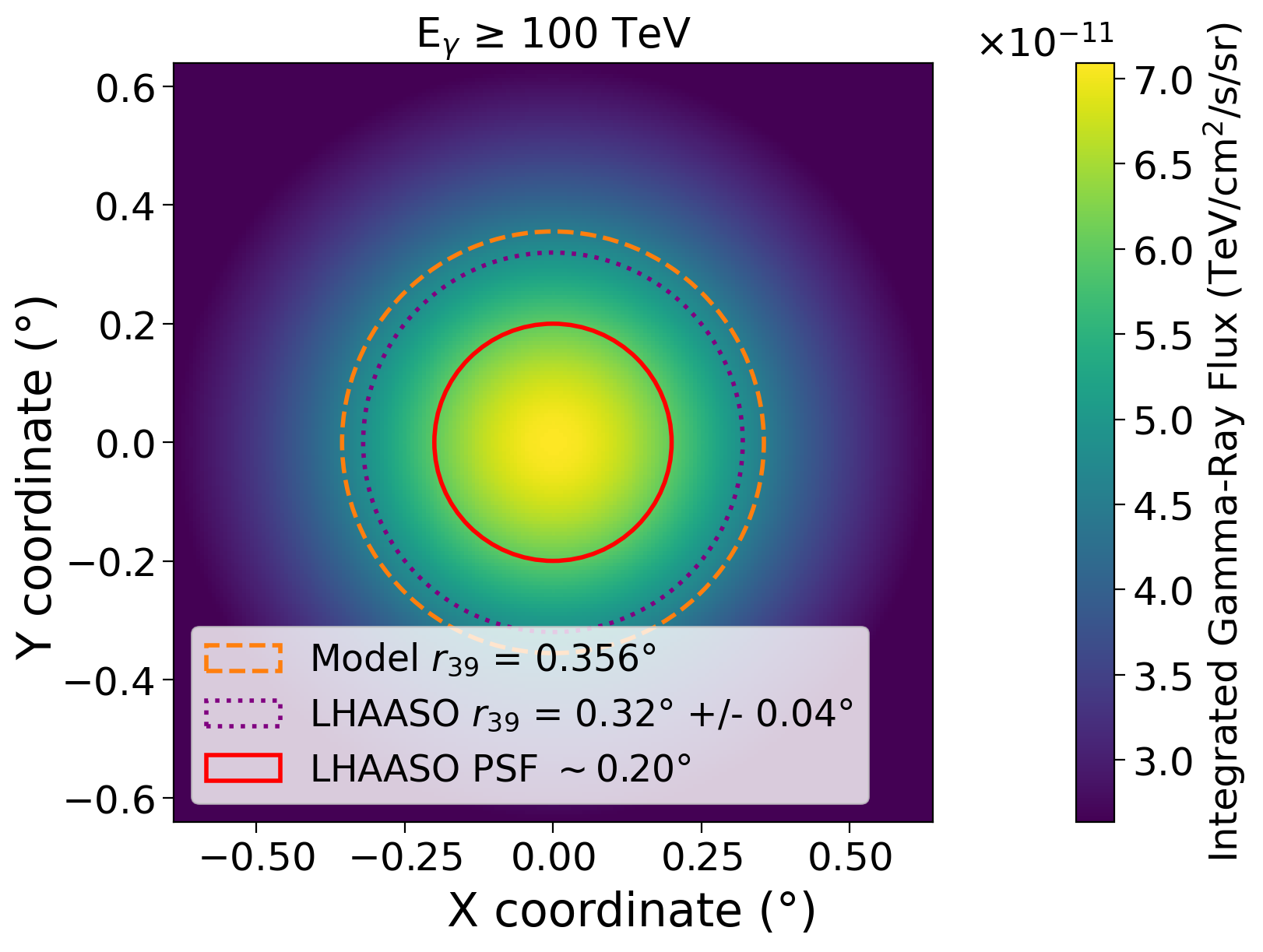}

\caption{The gamma-ray flux from the central source of the microquasar SS 433, as predicted by our leptonic model (with $\alpha=2.0$ and $E_c=1 \, {\rm PeV}$). LHASSO is not expected to be sensitive to the emission at $1-25$ or $25-100 \, {\rm TeV}$. In the lower frame, 39\% of the flux at $>100 \, {\rm TeV}$ is predicted to come from within a radius of $r_{39} = 0.36^{\circ}$, in good agreement with LHAASO's measurement of $r_{39} = 0.32^{\circ} \pm 0.04^{\circ}$. This concordance favors a leptonic origin for the ultrahigh-energy emission from this source. LHAASO's point spread function is taken from ~\cite{LHAASO:2024psv}.}
\label{fig:SS433_cent_map}
\end{figure}

In addition to the spectrum, we compare the angular distribution of the gamma rays observed by LHAASO with that predicted by our leptonic model. In Fig.~\ref{fig:SS433_cent_map}, we show the predicted gamma-ray flux per solid angle from this source, integrated over three energy bands, for model parameters $\alpha=2.0$ and $E_c=1 \, {\rm PeV}$. Comparing these results to the measurements shown in Fig.~1 of Ref.~\cite{LHAASO:2024psv}, we conclude that LHAASO would not be sensitive to the emission predicted at $1-25$ or $25-100$ TeV, as shown in the upper panels.

The LHAASO Collaboration reports that the emission from the central source is best described by a radially symmetric 2D Gaussian, with a width such that 39\% of the total flux above 100 TeV originates from a region of radius $r_{39} = 0.32^{\circ} \pm 0.04^{\circ}$. In the lower frame of Fig.~\ref{fig:SS433_cent_map}, 39\% of the total model flux comes from within $r_{39} = 0.36^{\circ}$, in good agreement with LHAASO's measurement. We emphasize that we did not fit any spatial parameters to achieve this agreement; rather, the simple leptonic model adopted in this study naturally predicts a degree of spatial extension that matches the data. The same would not be expected in hadronic models.

To summarize this section, we find that the spectral and morphological characteristics of SS 433's central source can be explained by the leptonic process of inverse Compton scattering. Previous claims that the $> 100 \, {\rm TeV}$ gamma-ray emission from this source could not be attributed to leptonic processes are only valid if the magnetic field in the region (of order 50 pc) surrounding the central source is taken to be much stronger than that found in the ISM.

\begin{figure}[t]
\centering
\includegraphics[width=1.0\linewidth]{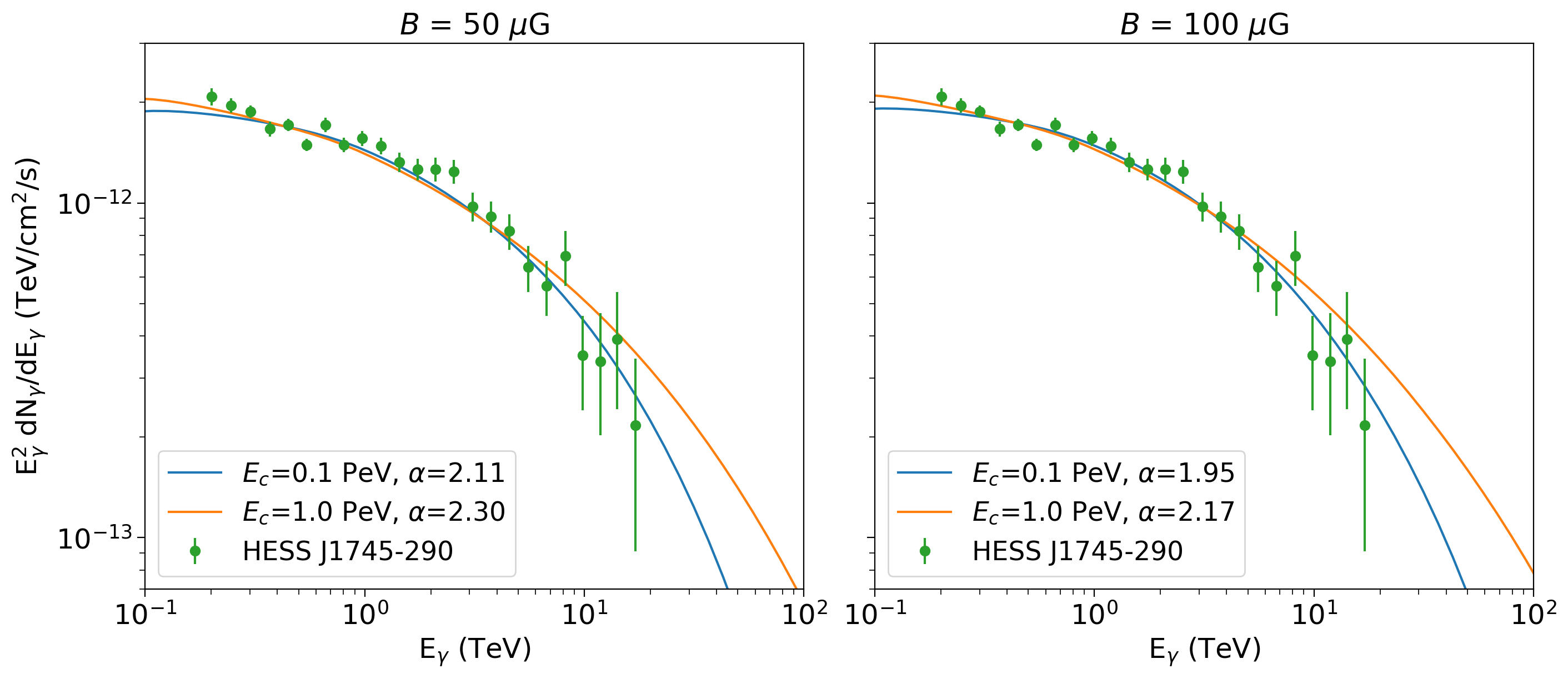}\\
\vspace{0.2cm}
\includegraphics[width=1.0\linewidth]{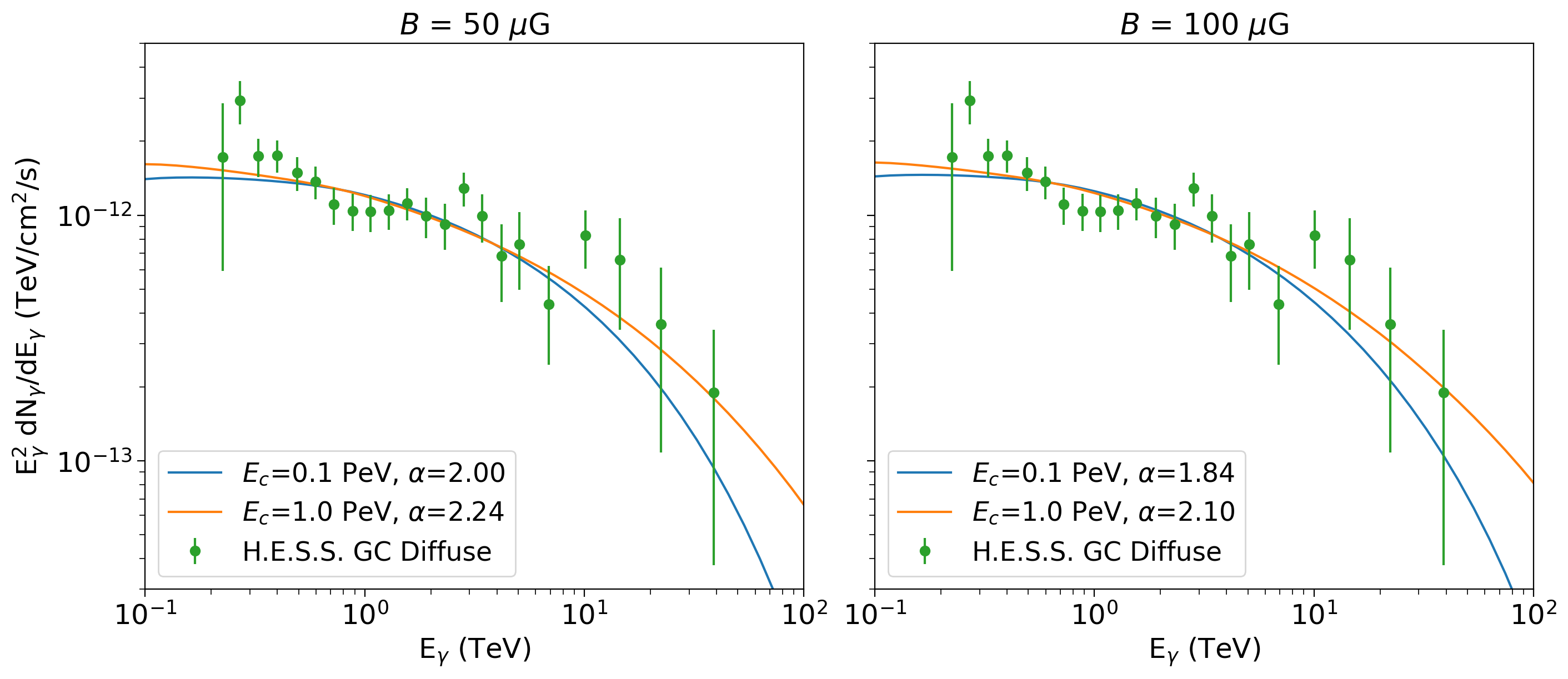}
\caption{The spectrum of the Galactic Center point source, HESS J1745-290 (upper frames), and of the surrounding diffuse emission (lower frames), compared to the predictions of our leptonic model. We have adopted a magnetic field strength of either $50\, \mu{\rm G}$ (left) or $100 \, \mu{\rm G}$ (right), chosen an electron cutoff energy of $E_c= 0.1 \, {\rm PeV}$ or 1 PeV, and adjusted the spectral index, $\alpha$, to obtain the best fit. Our simple leptonic model can achieve broad agreement with the spectrum observed by H.E.S.S.\label{fig:GC_PS_spectra}}
\end{figure}

\subsection{The Galactic Center} \label{GC}

The H.E.S.S. Collaboration has reported the detection of VHE emission from the Milky Way's Galactic Center (GC)~\cite{HESS:2016pst}. They separate this emission into the point-like source, HESS J1745-290, and a spatially extended component, both of which are centered at the GC. As we did in the case of SS 433, we use our leptonic model to fit the spectrum of HESS J1745-290 and of the surrounding diffuse emission. As a stronger magnetic field strength is expected to be present in the volume around the GC, we have fit these spectra adopting magnetic field strengths of $50\,\mu$G or $100\,\mu$G. We take the distance to the GC to be 8.5 kpc.  

The results of our fits to the GC point source, HESS J1745-290, are shown in the upper frames of Fig.~\ref{fig:GC_PS_spectra}, while those to the surrounding diffuse emission are presented in the lower frames. In these figures, we have adopted electron cutoff energies of $E_c = 0.1 \, {\rm PeV}$ and $1 \, {\rm PeV}$, and adjusted the spectral index to obtain the best fit. The comparison with the data demonstrates that this leptonic model can achieve broad agreement with the VHE gamma-ray spectrum that has been observed from the GC.

\begin{figure}[t]
    \centering
    \includegraphics[width=0.9\linewidth]{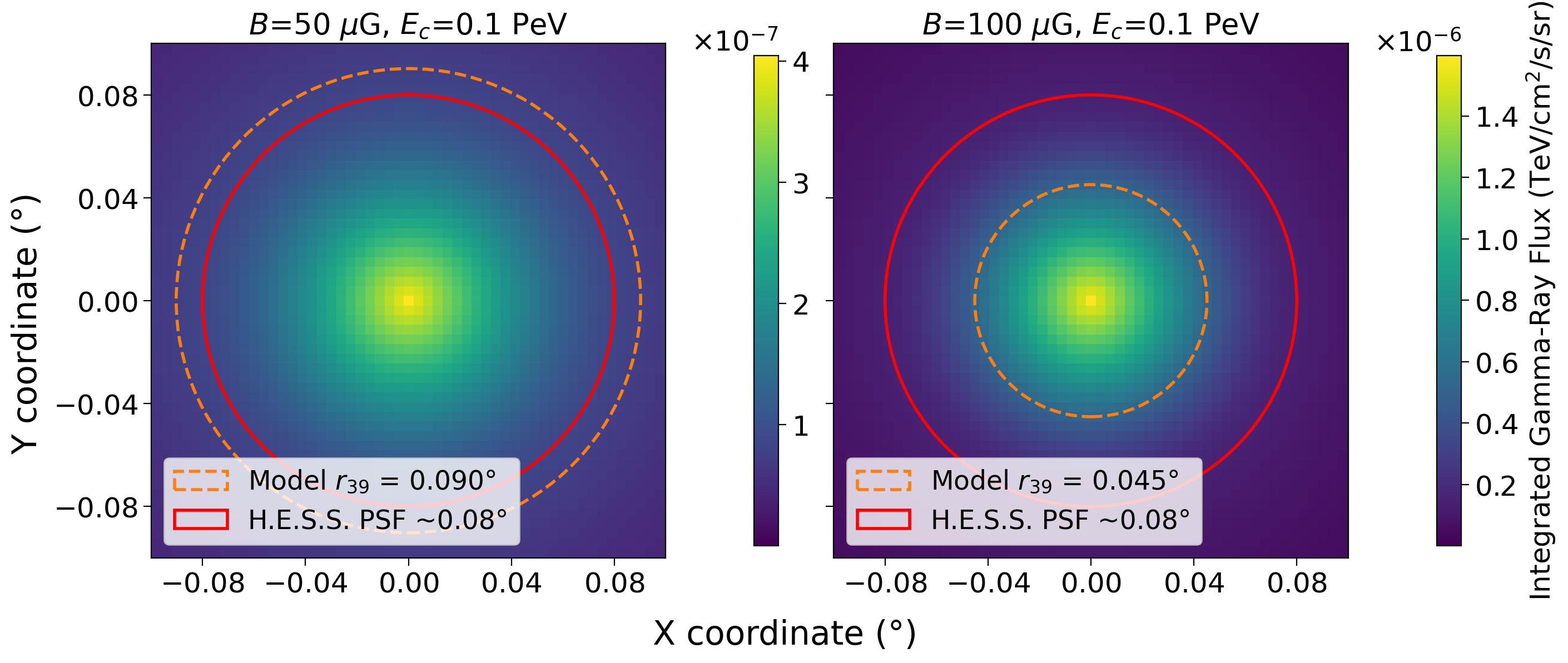}   
    \caption{The angular distribution of the gamma-ray flux, integrated above $1 \, {\rm TeV}$, predicted from electrons emitted from the Galactic Center. We show results for magnetic field strengths of $50 \, \mu{\rm G}$ (left) or $100 \, \mu{\rm G}$ (right), an electron cutoff energy of $E_c= 0.1 \, {\rm PeV}$, and have adjusted the electron spectral index to obtain the best fit to the observed spectrum of HESS J1745-290. In orange, we show the radius within which 39\% of the total flux is predicted to be emitted, and in red we show the approximate corresponding value of the H.E.S.S. point spread function at $10 \, {\rm TeV}$. This comparison demonstrates that the point-like nature of HESS J1745-290 is consistent with our leptonic model, as long as $B \gsim 50 \, \mu{\rm G}$ in the innermost $\sim 10 \, {\rm pc}$ of the Milky Way (or the diffusion coefficient is substantially smaller than that found in the ISM).}
    \label{fig:GC_PS_maps}
\end{figure}

In Fig.~\ref{fig:GC_PS_maps}, we plot the angular distribution of the gamma-ray emission, integrated above $1 \, {\rm TeV}$, predicted from a point source of electrons emitted from the Galactic Center, adopting a magnetic field of $50\, \mu{\rm G}$ (left) or $100\, \mu{\rm G}$ (right), an electron cutoff energy of $E_c= 0.1 \, {\rm PeV}$, and a spectral index of $\alpha=2.11$ (left) or $1.95$ (right). In orange, we show the radius within which 39\% of the total flux is predicted in our model, not including the effects of the H.E.S.S. point spread function (shown in red). From this comparison, we conclude that the point-like nature of HESS J1745-290 is consistent with our leptonic model, as long as $B \gsim 50 \, \mu{\rm G}$ in the innermost $\sim 10 \, {\rm pc}$ of the Milky Way (or the diffusion coefficient is substantially smaller than that found in the ISM).

The spatially extended emission reported by H.E.S.S. from the Galactic Center region includes a component that is roughly spherically symmetric with respect to the Galactic Center and a component that approximately traces the Galactic Plane~\cite{HESS:2016pst}. The former can be well fit by a population of leptonic sources with a number density that scales approximately as $n \propto 1/r^2$ (similar to the observed distribution of bulge stars~\cite{Coleman:2019kax}), assuming that the ratio of $\rho_{\rm rad}/\rho_{B}$ does not change much with distance to the Galactic Center. The latter component could be plausibly associated with leptonic sources within the Disk of the Milky Way. These sources could, for example, be a population of TeV halos, as suggested in Refs.~\cite{Hooper:2018sep,Hooper:2018fih}.

\subsection{TeV Halos}

VHE gamma-ray emission has been observed from the regions surrounding many young and middle-aged pulsars, and such TeV halos now appear to be approximately universal features of such systems~\cite{HAWC:2017kbo,HAWC:2019tcx,HESS:2017lee,Linden:2017vvb,Sudoh:2019lav,LHAASO:2021crt,Dekker:2023six}. The spectrum and morphology of TeV halos strongly suggest a leptonic origin. In particular, the gamma-ray emission observed from TeV halos can be straightforwardly accommodated if $\mathcal{O}(10\%)$ of the pulsar’s spindown power is converted into the acceleration of VHE electrons and positrons,
with a spectrum extending to several hundred TeV or higher. These sources feature a spectral break at an energy determined by the electron cooling timescale, which arises naturally in the context of these simple models.

Although the HAWC Collaboration has suggested that at least some TeV halos could be PeVatrons~\cite{HAWC:2019tcx}, it was shown in Ref.~\cite{Sudoh:2021} that a simple leptonic model could straightforwardly account for the observed spectra and morphologies of these sources~\cite{HAWC:2019tcx}. While we do not explicitly perform any calculations regarding TeV halos in this study, we direct the reader to Ref.~\cite{Sudoh:2021} for a detailed discussion of the UHE emission from TeV halos and the likely leptonic origin of that emission.

\begin{figure}[t]
\centering
\includegraphics[width=0.7\linewidth]{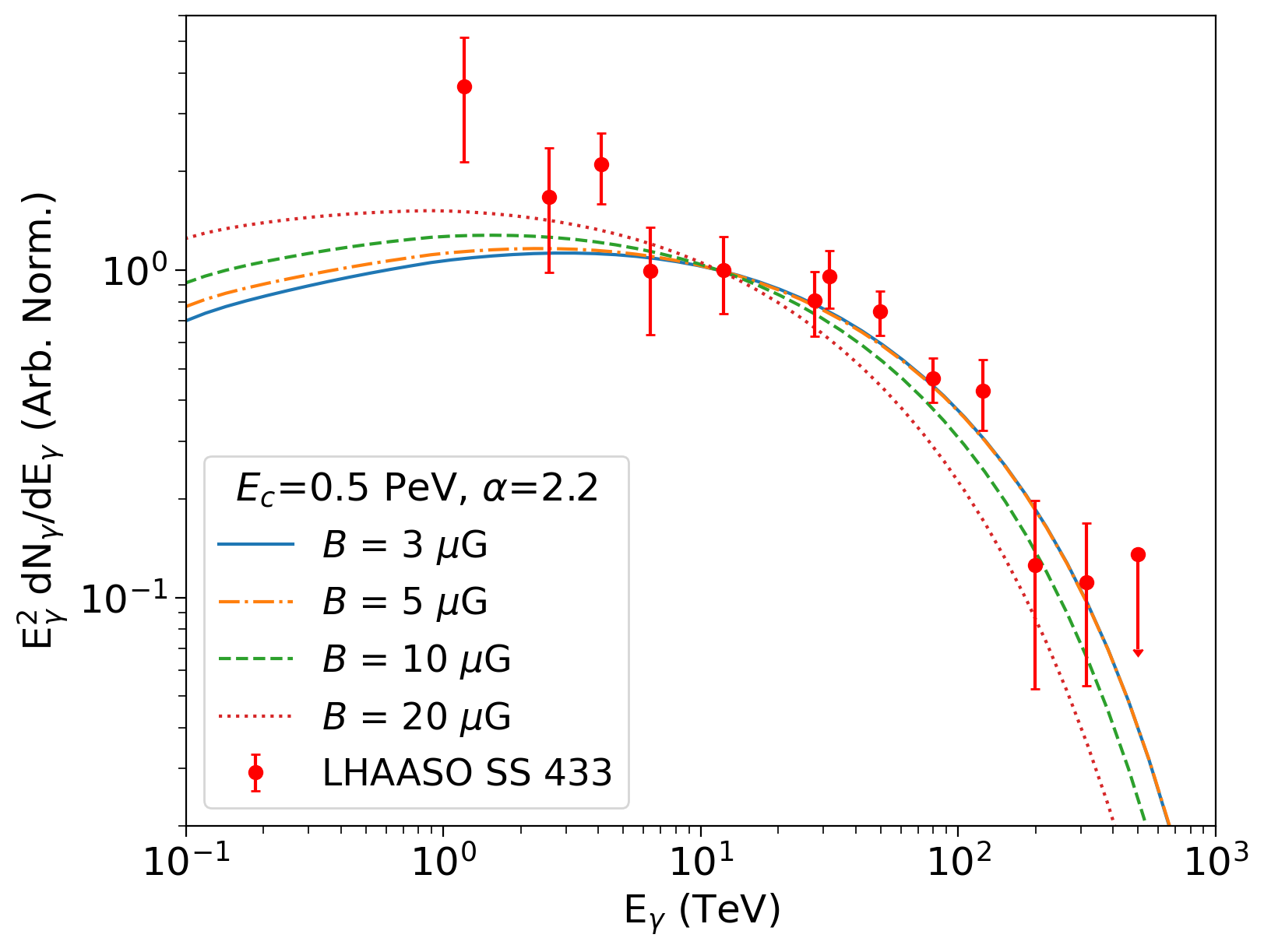}
\caption{The gamma-ray spectrum from a generic leptonic source, adopting an injected electron spectrum with a power-law index of 2.2 and an exponential cutoff at 0.5 PeV. The results are shown for several values of the magnetic field strength, and are compared to the spectrum of SS 433 (summing the central source and both lobes), as reported by the LHAASO Collaboration~\cite{LHAASO:2024psv}.   \label{fig:Shape}}
\end{figure}

\section{How to Identify a PeVatron} \label{shapes}

In this study, we have argued that the observation of $>100 \, {\rm TeV}$ emission from an astrophysical source does not necessarily imply a hadronic origin for those gamma rays, and that the leptonic process of inverse Compton scattering can accommodate such gamma rays, even after accounting for the effects of Klein-Nishina suppression. This, of course, does not mean that the sources of PeV-scale cosmic rays cannot be identified. Such a determination would, however, likely require additional information. This could include multiwavelength signatures, such as the combined observation of the pion bump and synchrotron peak~\cite{Fermi-LAT:2013iui,Yang:2018dsi}, or spatial correlation of the gamma-ray emission with gas~\cite{Aharonian:2000zd,Gabici:2006pg}. More distinctively, the observation of high-energy neutrinos from an astrophysical source would be a smoking gun for hadronic cosmic-ray acceleration~\cite{Halzen:2002pg,Learned:2000sw} (see, however, Ref~\cite{Hooper:2023ssc}).

In Fig.~\ref{fig:Shape}, we plot the spectral shape of the inverse Compton emission from a spectrum of electrons with a power-law index of $\alpha=2.2$ and an exponential cutoff at $E_c =  0.5 \, {\rm PeV}$. Results are shown for four values of the magnetic field strength. We compare this to the total spectrum observed from SS 433 by LHAASO. The figure demonstrates that the presence of $>100 \, {\rm TeV}$ gamma-ray emission from this source does not preclude a leptonic origin of this emission. In particular, for these parameter choices, the observed spectrum can be accommodated by a lepton source with these parameters as long as the magnetic field is not much stronger than a few microgauss (the argument in favor of hadronic acceleration in this source that was made in Ref.~\cite{LHAASO:2024psv} relied on the assumption of $B=20 \, \mu{\rm G}$). Similar figures corresponding to other parameter values can be found in the Appendix.

\begin{figure}[t]
\centering
\includegraphics[width=0.7\linewidth]{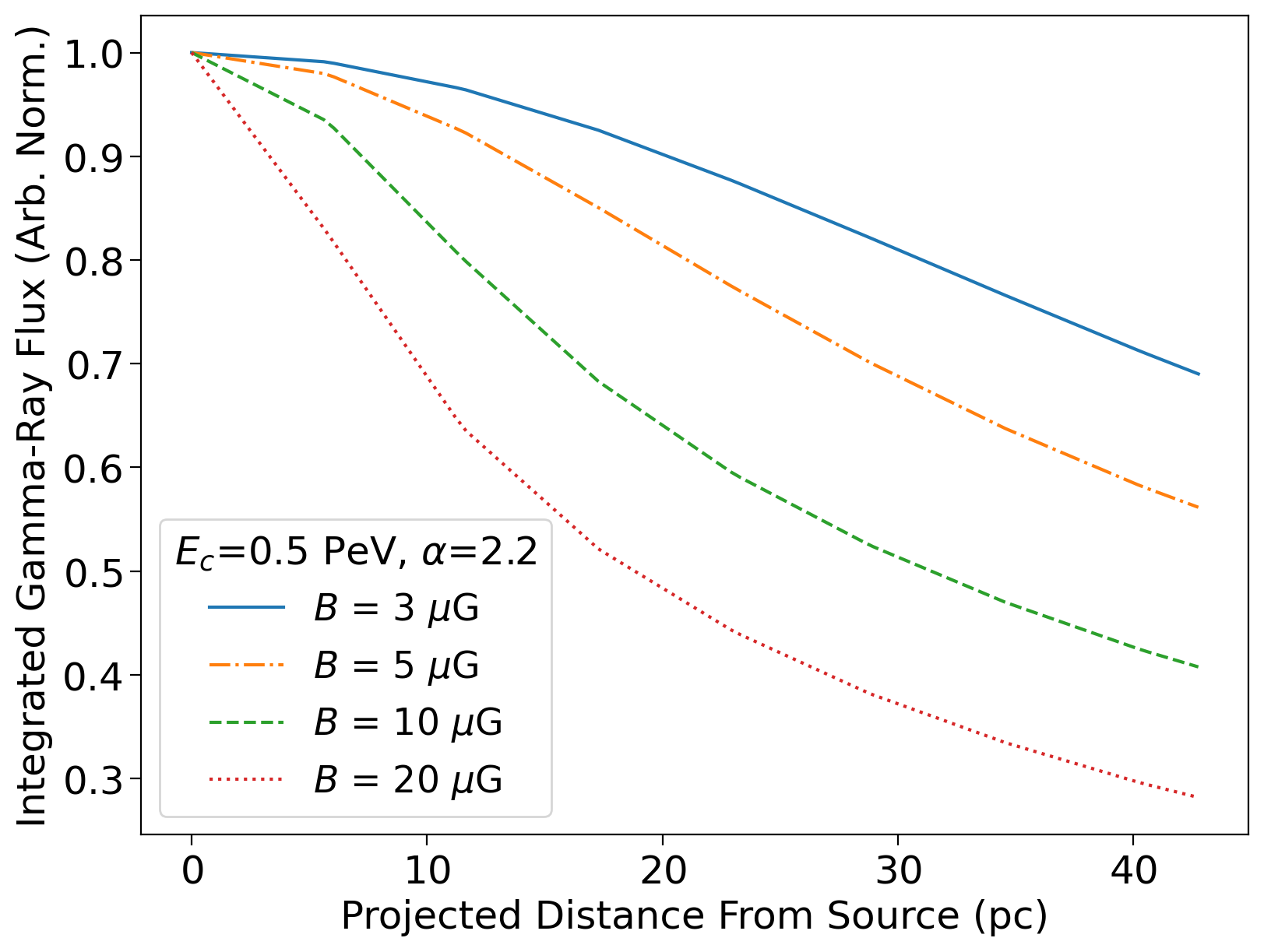}
\caption{The flux per solid angle of the VHE gamma-ray emission (integrated in the range $1-100 \, {\rm TeV}$) predicted from a leptonic source with an injected spectral index of $\alpha=2.2$, a cutoff energy of $0.5 \, {\rm PeV}$, and for four values of the magnetic field strength. Here we have assumed a standard (ISM-like) diffusion coefficient and radiation energy densities as described in \ref{leptonic_model}.}
\label{fig:radprofexample}
\end{figure}

The morphology of the gamma-ray emission from a source could also be used to discriminate between hadronic and leptonic models. In particular, if we assume standard (ISM-like) diffusion and energy loss rates, we can calculate the angular distribution of the gamma-ray emission that would be predicted from a leptonic source. In Fig.~\ref{fig:radprofexample}, we show the integrated gamma-ray flux per solid angle predicted from a leptonic source with an injected spectral index of $\alpha=2.2$, a cutoff energy of $0.5 \, {\rm PeV}$, and for three values of the magnetic field strength. A leptonic source with a profile of VHE gamma-ray emission that does not conform to this prediction must be in a region with a significantly different diffusion coefficient, radiation density, and/or magnetic field density. Alternatively, such a deviation could be interpreted as a signature of hadron acceleration. This conclusion would be further strengthened if the emission in question were observed to correlate with the surrounding gas density. Similar figures corresponding to other parameter values can be found in the Appendix.

\section{Summary and Conclusions} \label{conclusion}

It has long been suggested that the spectral feature in the cosmic-ray spectrum known as the ``knee'' could be associated with the maximum energy to which Galactic sources accelerate protons. This has motivated efforts to identify the sources of these PeV protons -- the Galactic PeVatrons. While there is evidence that at least some supernova remnants accelerate protons, such objects seem unlikely to reach PeV-scale energies. Microquasars, pulsar wind nebulae, star-forming regions, massive stars, and the Milky Way's central supermassive black hole have each been proposed as potential PeVatron candidates.

It has been argued that the observation of ultrahigh-energy gamma rays from an astrophysical source would constitute evidence that it is a PeVatron. In particular, inverse Compton scattering experiences Klein-Nishina suppression at PeV-scale energies, increasing the fraction of the total energy lost to synchrotron emission. Such arguments have been made in the context of the microquasar SS 433~\cite{LHAASO:2024psv}, the Milky Way's central supermassive black hole~\cite{HESS:2016pst}, the star-forming region Cygnus X~\cite{LHAASO:2023uhj}, and selected TeV halos~\cite{HAWC:2019tcx}. In this paper, we have argued that the observed characteristics of these sources can all be accommodated by leptonic models. More generally, the observation of PeV-scale gamma-rays from a source does not necessarily indicate that it accelerates hadronic cosmic rays.

In this paper, we have used a simple diffusion-loss model to predict the gamma-ray emission from accelerators of ultrahigh-energy electrons, and applied this model to several would-be PeVatrons. In each case, we found that the observed features of these sources could be explained by this leptonic scenario. In the case of the microquasar SS 433, our model predicts that the gamma-ray emission should have an angular extension that is in good agreement with that measured by LHAASO, favoring a leptonic origin for this signal.

For these reasons, the observation of ultrahigh-energy gamma-ray emission from a source is unlikely to be sufficient to convincingly demonstrate that it accelerates PeV-scale protons. A definitive identification of a PeVatron would require additional information, such as the combined observation of the pion bump and synchrotron peak or the spatial correlation of gamma-ray emission with gas. The most compelling signature of a PeVatron would, of course, be its production of $\gsim 100 \, {\rm TeV}$ neutrinos. Such a signal would constitute a smoking gun for PeV-scale hadronic cosmic-ray acceleration. Upcoming observatories, such as the Cherenkov Telescope Array Observatory (CTAO) and IceCube Gen-2, will be highly sensitive to these signatures and will be well positioned to discover the sources of the PeV-scale cosmic rays.

\bigskip

\section{Acknowledgements}
We would like to thank Ke Fang and Mukul Bhattacharya for valuable discussions.
ZCG and JV acknowledge funding from NSF awards 2413037 and 2013102. DH is supported by the Office of the Vice Chancellor for Research at the University of Wisconsin-Madison, with funding from the Wisconsin Alumni Research Foundation. In this work we have made use of the astropy, matplotlib, naima, numba, numpy, and scipy open source Python packages.

\section{Code Availability}

The code we have used for modeling the gamma-ray emission from a leptonic source has been made publicly available, along with example scripts for generating the spectrum, radial profile, and flux map of the predicted emission. See \url{https://github.com/zcurtisginsberg/Leptonic_Modeling_of_PeVatrons}

\bibliographystyle{JHEP}
\bibliography{biblio}

@article{Hooper:2023ssc,
    author = "Hooper, Dan and Plant, Kathryn",
    title = "{Leptonic Model for Neutrino Emission from Active Galactic Nuclei}",
    eprint = "2305.06375",
    archivePrefix = "arXiv",
    primaryClass = "astro-ph.HE",
    reportNumber = "FERMILAB-PUB-23-232-T",
    doi = "10.1103/PhysRevLett.131.231001",
    journal = "Phys. Rev. Lett.",
    volume = "131",
    number = "23",
    pages = "231001",
    year = "2023"
}

@article{Coleman:2019kax,
    author = "Coleman, Brenna and Paterson, Dylan and Gordon, Chris and Macias, Oscar and Ploeg, Harrison",
    title = "{Maximum Entropy Estimation of the Galactic Bulge Morphology via the VVV Red Clump}",
    eprint = "1911.04714",
    archivePrefix = "arXiv",
    primaryClass = "astro-ph.GA",
    doi = "10.1093/mnras/staa1281",
    journal = "Mon. Not. Roy. Astron. Soc.",
    volume = "495",
    number = "3",
    pages = "3350--3372",
    year = "2020"
}

@article{Arons:2012tz,
    author = "Arons, Jonathan",
    title = "{Pulsar Wind Nebulae as Cosmic Pevatrons: A Current Sheet's Tale}",
    eprint = "1208.5787",
    archivePrefix = "arXiv",
    primaryClass = "astro-ph.HE",
    doi = "10.1007/s11214-012-9885-1",
    journal = "Space Sci. Rev.",
    volume = "173",
    pages = "341--367",
    year = "2012"
}

@inproceedings{Porter:2005qx,
    author = "Porter, Troy A. and Strong, A. W.",
    title = "{A New estimate of the Galactic interstellar radiation field between 0.1 microns and 1000 microns}",
    booktitle = "{29th International Cosmic Ray Conference}",
    eprint = "astro-ph/0507119",
    archivePrefix = "arXiv",
    month = "7",
    year = "2005"
}

@article{Gaensler:2006ua,
    author = "Gaensler, Bryan M. and Slane, Patrick O.",
    title = "{The evolution and structure of pulsar wind nebulae}",
    eprint = "astro-ph/0601081",
    archivePrefix = "arXiv",
    doi = "10.1146/annurev.astro.44.051905.092528",
    journal = "Ann. Rev. Astron. Astrophys.",
    volume = "44",
    pages = "17--47",
    year = "2006"
}

@article{LHAASO:2025ysm,
    author = "Cao, Zhen and others",
    collaboration = "LHAASO",
    title = "{Cygnus X-3: A variable petaelectronvolt gamma-ray source}",
    eprint = "2512.16638",
    archivePrefix = "arXiv",
    primaryClass = "astro-ph.HE",
    month = "12",
    year = "2025"
}

@article{Brinkmann:2006zt,
    author = "Brinkmann, W. and Pratt, Gabriel W. and Rohr, S. and Kawai, N. and Burwitz, V.",
    title = "{XMM-Newton observations of the eastern jet of SS433}",
    eprint = "astro-ph/0610781",
    archivePrefix = "arXiv",
    doi = "10.1051/0004-6361:20065570",
    journal = "Astron. Astrophys.",
    volume = "463",
    pages = "611--620",
    year = "2007"
}

@article{Yang:2018dsi,
    author = "Yang, Rui-zhi and Kafexhiu, Ervin and Aharonian, Felix",
    title = "{On the shape of the gamma-ray spectrum around the ''$\pi^0$-bump''}",
    eprint = "1803.05072",
    archivePrefix = "arXiv",
    primaryClass = "astro-ph.HE",
    doi = "10.1051/0004-6361/201730908",
    journal = "Astron. Astrophys.",
    volume = "615",
    pages = "A108",
    year = "2018"
}

@article{Fermi-LAT:2013iui,
    author = "Ackermann, M. and others",
    collaboration = "Fermi-LAT",
    title = "{Detection of the Characteristic Pion-Decay Signature in Supernova Remnants}",
    eprint = "1302.3307",
    archivePrefix = "arXiv",
    primaryClass = "astro-ph.HE",
    doi = "10.1126/science.1231160",
    journal = "Science",
    volume = "339",
    pages = "807",
    year = "2013"
}

@article{Aharonian:2000zd,
    author = "Aharonian, F. A.",
    title = "{Gamma-rays from molecular clouds}",
    eprint = "astro-ph/0012290",
    archivePrefix = "arXiv",
    doi = "10.1023/A:1013845015364",
    journal = "Space Sci. Rev.",
    volume = "99",
    pages = "187",
    year = "2001"
}

@article{Gabici:2006pg,
    author = "Gabici, Stefano and Aharonian, Felix and Blasi, Pasquale",
    title = "{Gamma rays from molecular clouds}",
    eprint = "astro-ph/0610032",
    archivePrefix = "arXiv",
    doi = "10.1007/s10509-007-9427-6",
    journal = "Astrophys. Space Sci.",
    volume = "309",
    pages = "365--371",
    year = "2007"
}

@article{Halzen:2002pg,
    author = "Halzen, Francis and Hooper, Dan",
    title = "{High-energy neutrino astronomy: The cosmic ray connection}",
    eprint = "astro-ph/0204527",
    archivePrefix = "arXiv",
    doi = "10.1088/0034-4885/65/7/201",
    journal = "Rept. Prog. Phys.",
    volume = "65",
    pages = "1025--1078",
    year = "2002"
}

@article{Learned:2000sw,
    author = "Learned, J. G. and Mannheim, K.",
    title = "{High-energy neutrino astrophysics}",
    doi = "10.1146/annurev.nucl.50.1.679",
    journal = "Ann. Rev. Nucl. Part. Sci.",
    volume = "50",
    pages = "679--749",
    year = "2000"
}

@article{LHAASO:2021crt,
    author = "Aharonian, F. and others",
    collaboration = "LHAASO",
    title = "{Extended Very-High-Energy Gamma-Ray Emission Surrounding PSR J0622+3749 Observed by LHAASO-KM2A}",
    eprint = "2106.09396",
    archivePrefix = "arXiv",
    primaryClass = "astro-ph.HE",
    doi = "10.1103/PhysRevLett.126.241103",
    journal = "Phys. Rev. Lett.",
    volume = "126",
    number = "24",
    pages = "241103",
    year = "2021"
}

@article{Dekker:2023six,
    author = "Dekker, Ariane and Holst, Ian and Hooper, Dan and Leone, Giovani and Simon, Emily and Xiao, Huangyu",
    title = "{Diffuse ultrahigh-energy gamma-ray emission from TeV halos}",
    eprint = "2306.00051",
    archivePrefix = "arXiv",
    primaryClass = "astro-ph.HE",
    reportNumber = "FERMILAB-PUB-23-272-T",
    doi = "10.1103/PhysRevD.109.083026",
    journal = "Phys. Rev. D",
    volume = "109",
    number = "8",
    pages = "083026",
    year = "2024"
}

@article{HAWC:2017kbo,
    author = "Abeysekara, A. U. and others",
    collaboration = "HAWC",
    title = "{Extended gamma-ray sources around pulsars constrain the origin of the positron flux at Earth}",
    eprint = "1711.06223",
    archivePrefix = "arXiv",
    primaryClass = "astro-ph.HE",
    doi = "10.1126/science.aan4880",
    journal = "Science",
    volume = "358",
    number = "6365",
    pages = "911--914",
    year = "2017"
}

@article{HAWC:2019tcx,
    author = "Abeysekara, A. U. and others",
    collaboration = "HAWC",
    title = "{Multiple Galactic Sources with Emission Above 56 TeV Detected by HAWC}",
    eprint = "1909.08609",
    archivePrefix = "arXiv",
    primaryClass = "astro-ph.HE",
    doi = "10.1103/PhysRevLett.124.021102",
    journal = "Phys. Rev. Lett.",
    volume = "124",
    number = "2",
    pages = "021102",
    year = "2020"
}

@article{HESS:2017lee,
    author = "Abdalla, H. and others",
    collaboration = "HESS",
    title = "{The population of TeV pulsar wind nebulae in the H.E.S.S. Galactic Plane Survey}",
    eprint = "1702.08280",
    archivePrefix = "arXiv",
    primaryClass = "astro-ph.HE",
    doi = "10.1051/0004-6361/201629377",
    journal = "Astron. Astrophys.",
    volume = "612",
    pages = "A2",
    year = "2018"
}

@article{Linden:2017vvb,
    author = "Linden, Tim and Auchettl, Katie and Bramante, Joseph and Cholis, Ilias and Fang, Ke and Hooper, Dan and Karwal, Tanvi and Li, Shirley Weishi",
    title = "{Using HAWC to discover invisible pulsars}",
    eprint = "1703.09704",
    archivePrefix = "arXiv",
    primaryClass = "astro-ph.HE",
    reportNumber = "FERMILAB-PUB-17-080-A",
    doi = "10.1103/PhysRevD.96.103016",
    journal = "Phys. Rev. D",
    volume = "96",
    number = "10",
    pages = "103016",
    year = "2017"
}

@article{Sudoh:2019lav,
    author = "Sudoh, Takahiro and Linden, Tim and Beacom, John F.",
    title = "{TeV Halos are Everywhere: Prospects for New Discoveries}",
    eprint = "1902.08203",
    archivePrefix = "arXiv",
    primaryClass = "astro-ph.HE",
    doi = "10.1103/PhysRevD.100.043016",
    journal = "Phys. Rev. D",
    volume = "100",
    number = "4",
    pages = "043016",
    year = "2019"
}

@article{Hooper:2018fih,
    author = "Hooper, Dan and Linden, Tim",
    title = "{Millisecond Pulsars, TeV Halos, and Implications For The Galactic Center Gamma-Ray Excess}",
    eprint = "1803.08046",
    archivePrefix = "arXiv",
    primaryClass = "astro-ph.HE",
    reportNumber = "FERMILAB-PUB-18-069-A",
    doi = "10.1103/PhysRevD.98.043005",
    journal = "Phys. Rev. D",
    volume = "98",
    number = "4",
    pages = "043005",
    year = "2018"
}

@article{LHAASO:2024psv,
    author = "Cao, Zhen and others",
    collaboration = "LHAASO",
    title = "{Ultrahigh-energy gamma-ray emission associated with black hole{\textendash}jet systems}",
    eprint = "2410.08988",
    archivePrefix = "arXiv",
    primaryClass = "astro-ph.HE",
    doi = "10.1093/nsr/nwaf496",
    journal = "Natl. Sci. Rev.",
    volume = "12",
    number = "12",
    pages = "nwaf496",
    year = "2025"
}

@article{LHAASO:2023uhj,
    author = "Cao, Zhen and others",
    collaboration = "LHAASO",
    title = "{An ultrahigh-energy {\ensuremath{\gamma}}-ray bubble powered by a super PeVatron}",
    eprint = "2310.10100",
    archivePrefix = "arXiv",
    primaryClass = "astro-ph.HE",
    doi = "10.1016/j.scib.2023.12.040",
    journal = "Sci. Bull.",
    volume = "69",
    number = "4",
    pages = "449--457",
    year = "2024"
}

@article{Fujita:2016yvk,
    author = "Fujita, Yutaka and Murase, Kohta and Kimura, Shigeo S.",
    title = "{Sagittarius A* as an Origin of the Galactic PeV Cosmic Rays?}",
    eprint = "1604.00003",
    archivePrefix = "arXiv",
    primaryClass = "astro-ph.HE",
    reportNumber = "OU-TAP-384",
    doi = "10.1088/1475-7516/2017/04/037",
    journal = "JCAP",
    volume = "04",
    pages = "037",
    year = "2017"
}

@article{Guo:2016zjl,
    author = "Guo, Yi-Qing and Tian, Zhen and Wang, Zhen and Li, Hai-Jin and Chen, Tian-Lu",
    title = "{The Galactic Center: A Petaelectronvolt Cosmic-ray Acceleration Factory}",
    eprint = "1604.08301",
    archivePrefix = "arXiv",
    primaryClass = "astro-ph.HE",
    doi = "10.3847/1538-4357/aa5f58",
    journal = "Astrophys. J.",
    volume = "836",
    number = "2",
    pages = "233",
    year = "2017"
}

@article{HESS:2016pst,
    author = "Abramowski, A. and others",
    collaboration = "H.E.S.S.",
    title = "{Acceleration of petaelectronvolt protons in the Galactic Centre}",
    eprint = "1603.07730",
    archivePrefix = "arXiv",
    primaryClass = "astro-ph.HE",
    doi = "10.1038/nature17147",
    journal = "Nature",
    volume = "531",
    pages = "476",
    year = "2016"
}

@article{AGILE:2010egi,
    author = "Tavani, M. and others",
    collaboration = "AGILE",
    title = "{Direct Evidence for Hadronic Cosmic-Ray Acceleration in the Supernova Renmant IC 443}",
    eprint = "1001.5150",
    archivePrefix = "arXiv",
    primaryClass = "astro-ph.HE",
    doi = "10.1088/2041-8205/710/2/L151",
    journal = "Astrophys. J. Lett.",
    volume = "710",
    pages = "L151--L155",
    year = "2010"
}

@article{Kaci:2025gyb,
    author = "Kaci, Samy and Giacinti, Gwenael and Aharonian, Felix and Wang, Jie-Shuang",
    title = "{Microquasars as the major contributors to Galactic cosmic rays around the ''knee''}",
    eprint = "2510.01369",
    archivePrefix = "arXiv",
    primaryClass = "astro-ph.HE",
    month = "10",
    year = "2025"
}

@article{Jouvin:2017bwo,
    author = "Jouvin, L. and Lemi{\`e}re, A. and Terrier, R.",
    title = "{Does the SN rate explain the very high energy cosmic rays in the central 200 pc of our Galaxy?}",
    eprint = "1703.10398",
    archivePrefix = "arXiv",
    primaryClass = "astro-ph.HE",
    doi = "10.1093/mnras/stx361",
    journal = "Mon. Not. Roy. Astron. Soc.",
    volume = "467",
    number = "4",
    pages = "4622--4630",
    year = "2017"
}

@article{Aharonian:2018oau,
    author = "Aharonian, Felix and Yang, Ruizhi and de O{\~n}a Wilhelmi, Emma",
    title = "{Massive Stars as Major Factories of Galactic Cosmic Rays}",
    eprint = "1804.02331",
    archivePrefix = "arXiv",
    primaryClass = "astro-ph.HE",
    doi = "10.1038/s41550-019-0724-0",
    journal = "Nature Astron.",
    volume = "3",
    number = "6",
    pages = "561--567",
    year = "2019"
}

@article{Bykov:2020zqf,
    author = "Bykov, Andrei M. and Marcowith, Alexandre and Amato, Elena and Kalyashova, Maria E. and Kruijssen, J. M. Diederik and Waxman, Eli",
    title = "{High-Energy Particles and Radiation in Star-Forming Regions}",
    eprint = "2003.11534",
    archivePrefix = "arXiv",
    primaryClass = "astro-ph.HE",
    doi = "10.1007/s11214-020-00663-0",
    journal = "Space Sci. Rev.",
    volume = "216",
    number = "3",
    pages = "42",
    year = "2020"
}

@article{Lagage:1983zz,
    author = "Lagage, P. O. and Cesarsky, C. J.",
    title = "{The maximum energy of cosmic rays accelerated by supernova shocks}",
    journal = "Astron. Astrophys.",
    volume = "125",
    pages = "249--257",
    year = "1983"
}

@article{Cristofari:2020mdf,
    author = "Cristofari, Pierre and Blasi, Pasquale and Amato, Elena",
    title = "{The low rate of Galactic pevatrons}",
    eprint = "2007.04294",
    archivePrefix = "arXiv",
    primaryClass = "astro-ph.HE",
    doi = "10.1016/j.astropartphys.2020.102492",
    journal = "Astropart. Phys.",
    volume = "123",
    pages = "102492",
    year = "2020"
}

@article{Gabici:2019jvz,
    author = "Gabici, Stefano and Evoli, Carmelo and Gaggero, Daniele and Lipari, Paolo and Mertsch, Philipp and Orlando, Elena and Strong, Andrew and Vittino, Andrea",
    title = "{The origin of Galactic cosmic rays: challenges to the standard paradigm}",
    eprint = "1903.11584",
    archivePrefix = "arXiv",
    primaryClass = "astro-ph.HE",
    reportNumber = "Preprint: TTK-19-12",
    doi = "10.1142/S0218271819300222",
    journal = "Int. J. Mod. Phys. D",
    volume = "28",
    number = "15",
    pages = "1930022",
    year = "2019"
}

@article{Gabici:2016fup,
    author = "Gabici, Stefano",
    title = "{Gamma-Ray Emission from Supernova Remnants and Surrounding Molecular Clouds}",
    eprint = "1610.06234",
    archivePrefix = "arXiv",
    primaryClass = "astro-ph.HE",
    doi = "10.1063/1.4968887",
    journal = "AIP Conf. Proc.",
    volume = "1792",
    number = "1",
    pages = "020002",
    year = "2017"
}

@article{Bell:2013kq,
    author = "Bell, AR and Schure, KM and Reville, B and Giacinti, G",
    title = "{Cosmic ray acceleration and escape from supernova remnants}",
    eprint = "1301.7264",
    archivePrefix = "arXiv",
    primaryClass = "astro-ph.HE",
    doi = "10.1093/mnras/stt179",
    journal = "Mon. Not. Roy. Astron. Soc.",
    volume = "431",
    pages = "415",
    year = "2013"
}

@article{MAGIC:2017hoy,
    author = "Ahnen, M. L. and others",
    collaboration = "MAGIC",
    title = "{A cut-off in the TeV gamma-ray spectrum of the SNR Cassiopeia A}",
    eprint = "1707.01583",
    archivePrefix = "arXiv",
    primaryClass = "astro-ph.HE",
    doi = "10.1093/mnras/stx2079",
    journal = "Mon. Not. Roy. Astron. Soc.",
    volume = "472",
    number = "3",
    pages = "2956--2962",
    year = "2017",
    note = "[Erratum: Mon.Not.Roy.Astron.Soc. 476, 2874--2875 (2018)]"
}

@article{VERITAS:2017qhy,
    author = "Archambault, S. and others",
    collaboration = "VERITAS",
    title = "{Gamma-Ray Observations of Tycho{\textquoteright}s Supernova Remnant with VERITAS and Fermi}",
    eprint = "1701.06740",
    archivePrefix = "arXiv",
    primaryClass = "astro-ph.HE",
    doi = "10.3847/1538-4357/836/1/23",
    journal = "Astrophys. J.",
    volume = "836",
    number = "1",
    pages = "23",
    year = "2017"
}

@article{Hillas:2005cs,
    author = "Hillas, A. M.",
    title = "{Can diffusive shock acceleration in supernova remnants account for high-energy galactic cosmic rays?}",
    doi = "10.1088/0954-3899/31/5/R02",
    journal = "J. Phys. G",
    volume = "31",
    pages = "R95--R131",
    year = "2005"
}

@article{Stanev:1993tx,
    author = "Stanev, Todor and Biermann, Peter L. and Gaisser, Thomas K.",
    title = "{Cosmic rays. 4. The Spectrum and chemical composition above 10**4-GeV}",
    eprint = "astro-ph/9303006",
    archivePrefix = "arXiv",
    reportNumber = "MPIFR-518",
    journal = "Astron. Astrophys.",
    volume = "274",
    pages = "902",
    year = "1993"
}

@article{Kobayakawa:2000nq,
    author = "Kobayakawa, K. and Sato, Y. and Samura, T.",
    title = "{Acceleration of particles by oblique shocks and cosmic ray spectra around the knee region}",
    eprint = "astro-ph/0008209",
    archivePrefix = "arXiv",
    doi = "10.1103/PhysRevD.66.083004",
    journal = "Phys. Rev. D",
    volume = "66",
    pages = "083004",
    year = "2002"
}

@article{Apel:2013uni,
    author = "Apel, W. D. and others",
    title = "{KASCADE-Grande measurements of energy spectra for elemental groups of cosmic rays}",
    eprint = "1306.6283",
    archivePrefix = "arXiv",
    primaryClass = "astro-ph.HE",
    doi = "10.1016/j.astropartphys.2013.06.004",
    journal = "Astropart. Phys.",
    volume = "47",
    pages = "54--66",
    year = "2013"
}

@article{LHAASO:2024knt,
    author = "Cao, Zhen and others",
    collaboration = "LHAASO",
    title = "{Measurements of All-Particle Energy Spectrum and Mean Logarithmic Mass of Cosmic Rays from 0.3 to 30~PeV with LHAASO-KM2A}",
    eprint = "2403.10010",
    archivePrefix = "arXiv",
    primaryClass = "astro-ph.HE",
    doi = "10.1103/PhysRevLett.132.131002",
    journal = "Phys. Rev. Lett.",
    volume = "132",
    number = "13",
    pages = "131002",
    year = "2024"
}

@article{EAS-TOP:2004aim,
    author = "Aglietta, M. and others",
    collaboration = "EAS-TOP",
    title = "{The cosmic ray primary composition in the 'knee' region through the EAS electromagnetic and muon measurements at EAS-TOP}",
    doi = "10.1016/j.astropartphys.2004.04.005",
    journal = "Astropart. Phys.",
    volume = "21",
    pages = "583--596",
    year = "2004"
}

@article{KASCADE:2005ynk,
    author = "Antoni, T. and others",
    collaboration = "KASCADE",
    title = "{KASCADE measurements of energy spectra for elemental groups of cosmic rays: Results and open problems}",
    eprint = "astro-ph/0505413",
    archivePrefix = "arXiv",
    doi = "10.1016/j.astropartphys.2005.04.001",
    journal = "Astropart. Phys.",
    volume = "24",
    pages = "1--25",
    year = "2005"
}

@article{IceCube:2012vv,
    author = "Abbasi, R. and others",
    collaboration = "IceCube",
    title = "{Cosmic Ray Composition and Energy Spectrum from 1-30 PeV Using the 40-String Configuration of IceTop and IceCube}",
    eprint = "1207.3455",
    archivePrefix = "arXiv",
    primaryClass = "astro-ph.HE",
    doi = "10.1016/j.astropartphys.2012.11.003",
    journal = "Astropart. Phys.",
    volume = "42",
    pages = "15--32",
    year = "2013"
}

@article{Hoerandel:2004gv,
    author = "Hoerandel, Jorg R.",
    title = "{Models of the knee in the energy spectrum of cosmic rays}",
    eprint = "astro-ph/0402356",
    archivePrefix = "arXiv",
    doi = "10.1016/j.astropartphys.2004.01.004",
    journal = "Astropart. Phys.",
    volume = "21",
    pages = "241--265",
    year = "2004"
}

@article{Allard:2005ha,
    author = "Allard, D. and Parizot, Etienne and Khan, E. and Goriely, S. and Olinto, A. V.",
    title = "{UHE nuclei propagation and the interpretation of the ankle in the cosmic-ray spectrum}",
    eprint = "astro-ph/0505566",
    archivePrefix = "arXiv",
    doi = "10.1051/0004-6361:200500199",
    journal = "Astron. Astrophys.",
    volume = "443",
    pages = "L29--L32",
    year = "2005"
}

@article{Kachelriess:2019oqu,
    author = "Kachelriess, M. and Semikoz, D. V.",
    title = "{Cosmic Ray Models}",
    eprint = "1904.08160",
    archivePrefix = "arXiv",
    primaryClass = "astro-ph.HE",
    doi = "10.1016/j.ppnp.2019.07.002",
    journal = "Prog. Part. Nucl. Phys.",
    volume = "109",
    pages = "103710",
    year = "2019"
}

@article{Blumenthal:1970gc,
    author = "Blumenthal, G. R. and Gould, R. J.",
    title = "{Bremsstrahlung, synchrotron radiation, and Compton scattering of high-energy electrons traversing dilute gases}",
    doi = "10.1103/RevModPhys.42.237",
    journal = "Rev. Mod. Phys.",
    volume = "42",
    pages = "237--270",
    year = "1970"
}

@article{Porter:2021tlr,
    author = "Porter, Troy A. and Johannesson, Gudlaugur and Moskalenko, V. Igor",
    title = "{The GALPROP Cosmic-ray Propagation and Nonthermal Emissions Framework: Release v57}",
    eprint = "2112.12745",
    archivePrefix = "arXiv",
    primaryClass = "astro-ph.HE",
    doi = "10.3847/1538-4365/ac80f6",
    journal = "Astrophys. J. Supp.",
    volume = "262",
    number = "1",
    pages = "30",
    year = "2022"
}

@article{Zabalza:2015bsa,
    author = "Zabalza, V{\'\i}ctor",
    title = "{Naima: a Python package for inference of relativistic particle energy distributions from observed nonthermal spectra}",
    eprint = "1509.03319",
    archivePrefix = "arXiv",
    primaryClass = "astro-ph.HE",
    doi = "10.22323/1.236.0922",
    journal = "PoS",
    volume = "ICRC2015",
    pages = "922",
    year = "2016"
}

@article{Khangulyan:2013hwa,
    author = "Khangulyan, Dmitry and Aharonian, Felix A. and Kelner, Stanislav R.",
    title = "{Simple Analytical Approximations for Treatment of Inverse Compton Scattering of Relativistic Electrons in the Blackbody Radiation Field}",
    eprint = "1310.7971",
    archivePrefix = "arXiv",
    primaryClass = "astro-ph.HE",
    doi = "10.1088/0004-637X/783/2/100",
    journal = "Astrophys. J.",
    volume = "783",
    pages = "100",
    year = "2014"
}

@article{Cherepashchuk:2019ibu,
    author = "Cherepashchuk, Anatol and Postnov, Konstantin and Molkov, Sergey and Antokhina, Eleonora and Belinski, Alexander",
    title = "{SS433: A massive X-ray binary in an advanced evolutionary stage}",
    eprint = "1905.02938",
    archivePrefix = "arXiv",
    primaryClass = "astro-ph.HE",
    doi = "10.1016/j.newar.2020.101542",
    journal = "New Astron. Rev.",
    volume = "89",
    pages = "101542",
    year = "2020"
}

@article{Stephenson:1977apr,
       author = "Stephenson, C.B. and Sanduleak, N.",
        title = "{New H-alpha emission stars in the Milky Way.}",
      journal = "The Astrophysical Journal Supplement Series",
         year = 1977,
        month = apr,
       volume = "33",
        pages = "459-469",
          doi = "10.1086/190437",
}

@article{HAWC:2018gwz,
    author = "Abeysekara, A. U. and others",
    collaboration = "HAWC",
    title = "{Very high energy particle acceleration powered by the jets of the microquasar SS 433}",
    eprint = "1810.01892",
    archivePrefix = "arXiv",
    primaryClass = "astro-ph.HE",
    doi = "10.1038/s41586-018-0565-5",
    journal = "Nature",
    volume = "562",
    number = "7725",
    pages = "82--85",
    year = "2018",
    note = "[Erratum: Nature 564, E38 (2018)]"
}

@article{HAWC:2024ysp,
    author = "Alfaro, R. and others",
    collaboration = "HAWC",
    title = "{Spectral Study of Very-high-energy Gamma Rays from SS 433 with HAWC}",
    eprint = "2410.21796",
    archivePrefix = "arXiv",
    primaryClass = "astro-ph.HE",
    doi = "10.3847/1538-4357/ad7e1b",
    journal = "Astrophys. J.",
    volume = "976",
    number = "1",
    pages = "30",
    year = "2024"
}

@article{HESS:2024jan,
       author = {{H.~E.~S.~S. Collaboration}},
        title = "{Acceleration and transport of relativistic electrons in the jets of the microquasar SS 433}",
      journal = {Science},
         year = 2024,
        month = jan,
       volume = {383},
       number = {6681},
        pages = {402-406},
          doi = {10.1126/science.adi2048},
archivePrefix = {arXiv},
       eprint = {2401.16019},
 primaryClass = {astro-ph.HE},
       adsurl = {https://ui.adsabs.harvard.edu/abs/2024Sci...383..402H}
}

@article{Klein:2025sep,
       author = {{Kleiner}, Tobias and {for the VERITAS Collaboration}},
        title = "{VERITAS Observations of the Microquasar SS 433}",
      journal = {arXiv e-prints},
         year = 2025,
        month = sep,
          eid = {arXiv:2509.21063},
        pages = {arXiv:2509.21063},
          doi = {10.48550/arXiv.2509.21063},
archivePrefix = {arXiv},
       eprint = {2509.21063},
 primaryClass = {astro-ph.HE},
       adsurl = {https://ui.adsabs.harvard.edu/abs/2025arXiv250921063K}
}

@article{Zealey:1980sep,
    author = "Zealey, W.J. and Dopita, M.A. and Malin, D.F.",
    title = "{The interaction between the relativistic jets of SS433 and the interstellar medium}",
    doi = "10.1093/mnras/192.4.731",
    journal = "Mon. Not. Roy. Astron. Soc.",
    volume = "192",
    number = "4",
    pages = "731--734",
    year = "1980"
}

@article{Goodall:2011jul,
    author = "Goodall, Paul T. and Alouani-Bibi, Fathallah and Blundell, Katherine M.",
    title = "{When microquasar jets and supernova collide: hydrodynamically simulating the SS 433–W 50 interaction}",
    doi = "10.1111/j.1365-2966.2011.18388.x",
    journal = "Mon. Not. Roy. Astron. Soc.",
    volume = "414",
    number = "4",
    pages = "2838--2859",
    year = "2011"
}

@article{Han:2020jun,
       author = "Han, Qin and Li, Xiang-Dong",
        title = "{On the Formation of SS433}",
      journal = "The Astrophysical Journal",
         year = 2020,
        month = jun,
       volume = "896",
       number = "1",
          eid = "34",
        pages = "34",
          doi = "10.3847/1538-4357/ab8d3d",
archivePrefix = "arXiv",
       eprint = "2004.12547",
 primaryClass = "astro-ph.HE"
}

@article{Su:2018aug,
       author = "Su, Yang and Zhou, Xin and Yang, Ji and Chen, Yang and Chen, Xuepeng and Zhang, Shaobo",
        title = "{The Large-scale Interstellar Medium of SS 433/W50 Revisited}",
      journal = "The Astrophysical Journal",
         year = 2018,
        month = aug,
       volume = "863",
       number = "1",
          eid = "103",
        pages = "103",
          doi = "10.3847/1538-4357/aad04e",
archivePrefix = "arXiv",
       eprint = "1807.03737",
 primaryClass = "astro-ph.GA"
}

@misc{Carpio:2025,
      title="{Multimessenger Emission from Very-High-Energy Black Hole-Jet Systems in the Milky Way}", 
      author={Jose Carpio and Ali Kheirandish and Bing Zhang},
      year={2025},
      eprint={2506.22550},
      archivePrefix={arXiv},
      primaryClass={astro-ph.HE},
      url={https://arxiv.org/abs/2506.22550}, 
}

@misc{Basanti:2025,
      title={Hadronic Emissions from the Microquasar {V4641} {Sgr}, {SS433}, and its implications in the diffuse {Galactic} emission}, 
      author={Basanti Paul and Abhijit Roy and Jagdish C. Joshi and Debanjan Bose},
      year={2025},
      eprint={2512.05839},
      archivePrefix={arXiv},
      primaryClass={astro-ph.HE},
      url={https://arxiv.org/abs/2512.05839}, 
}

@article{Ohira:2024qtr,
    author = "Ohira, Yutaka",
    title = "{Very-high-energy gamma-rays from cosmic rays escaping from Galactic black hole binaries}",
    eprint = "2410.22976",
    archivePrefix = "arXiv",
    primaryClass = "astro-ph.HE",
    doi = "10.1093/mnras/staf1176",
    journal = "Mon. Not. Roy. Astron. Soc.",
    volume = "541",
    number = "3",
    pages = "2434--2439",
    year = "2025"
}

@article{Sudoh:2021,
   title={The highest energy {HAWC} sources are likely leptonic and powered by pulsars},
   volume={2021},
   ISSN={1475-7516},
   url={http://dx.doi.org/10.1088/1475-7516/2021/08/010},
   DOI={10.1088/1475-7516/2021/08/010},
   number={08},
   journal={Journal of Cosmology and Astroparticle Physics},
   publisher={IOP Publishing},
   author={Sudoh, Takahiro and Linden, Tim and Hooper, Dan},
   year={2021},
   month=aug, pages={010} }

@article{Hooper:2018sep,
       author = {{Hooper}, Dan and {Cholis}, Ilias and {Linden}, Tim},
        title = "{TeV gamma rays from Galactic Center pulsars}",
      journal = {Physics of the Dark Universe},
         year = 2018,
        month = sep,
       volume = {21},
          eid = {40},
        pages = {40},
          doi = {10.1016/j.dark.2018.05.004},
archivePrefix = {arXiv},
       eprint = {1705.09293},
 primaryClass = {astro-ph.HE},
       adsurl = {https://ui.adsabs.harvard.edu/abs/2018PDU....21...40H}
}

\section*{Appendix: The Spectrum and Angular Distribution of Inverse Compton Emission From a Generic Leptonic Accelerator} \label{appendix_B}

This Appendix contains results similar to those presented in Figs.~\ref{fig:Shape} and~\ref{fig:radprofexample}, but adopting other parameter values for the injected spectrum of VHE electrons.

\begin{figure}[h]
\centering
\includegraphics[width=1.0\linewidth]{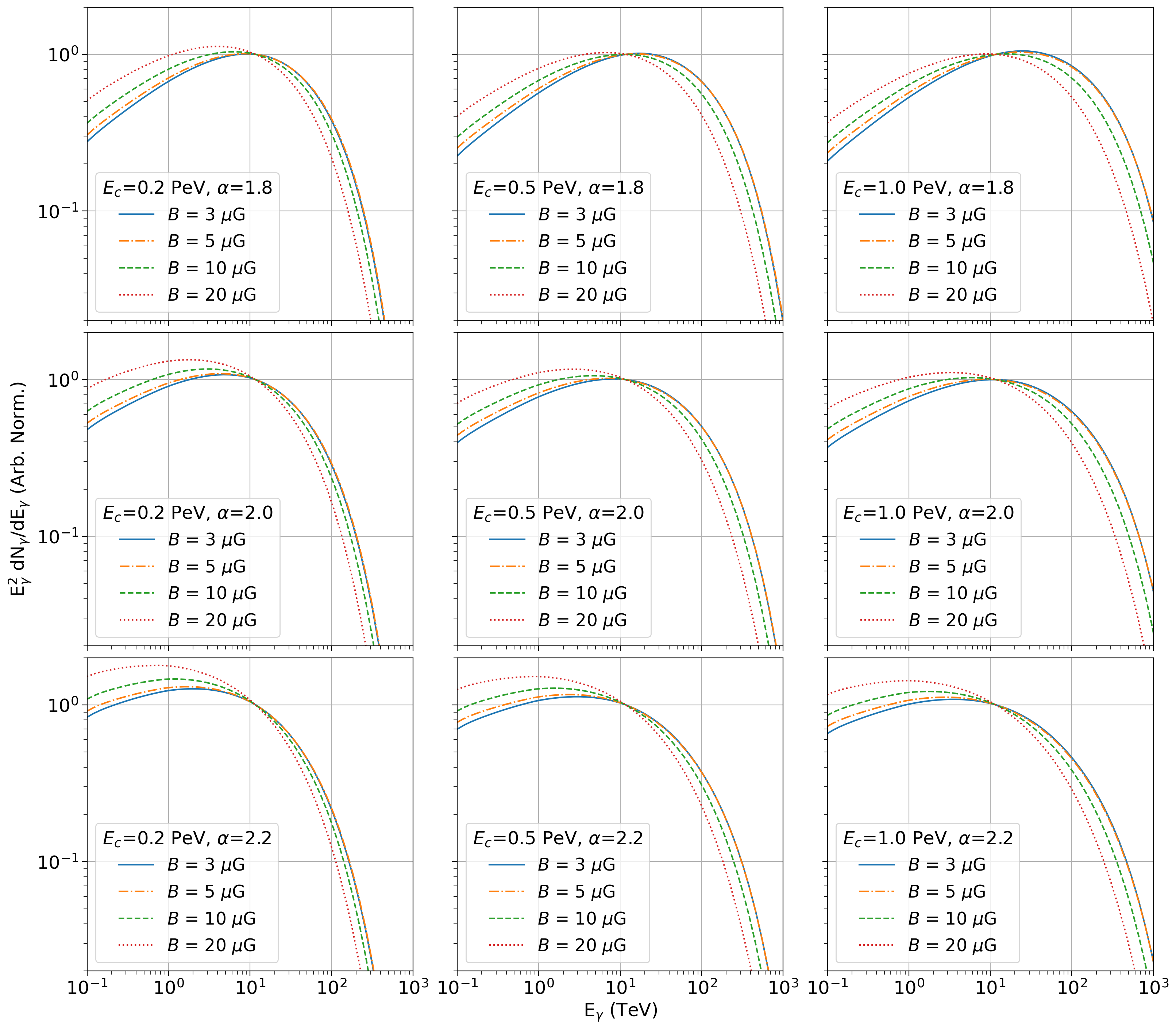}
\caption{As in Fig.~\ref{fig:Shape}, but adopting other values for the spectral index, $\alpha$, and exponential cutoff, $E_c$, of the injected electrons.}
\label{fig:shape01}
\end{figure}

\begin{figure}[h]
\centering
\includegraphics[width=1.0\linewidth]{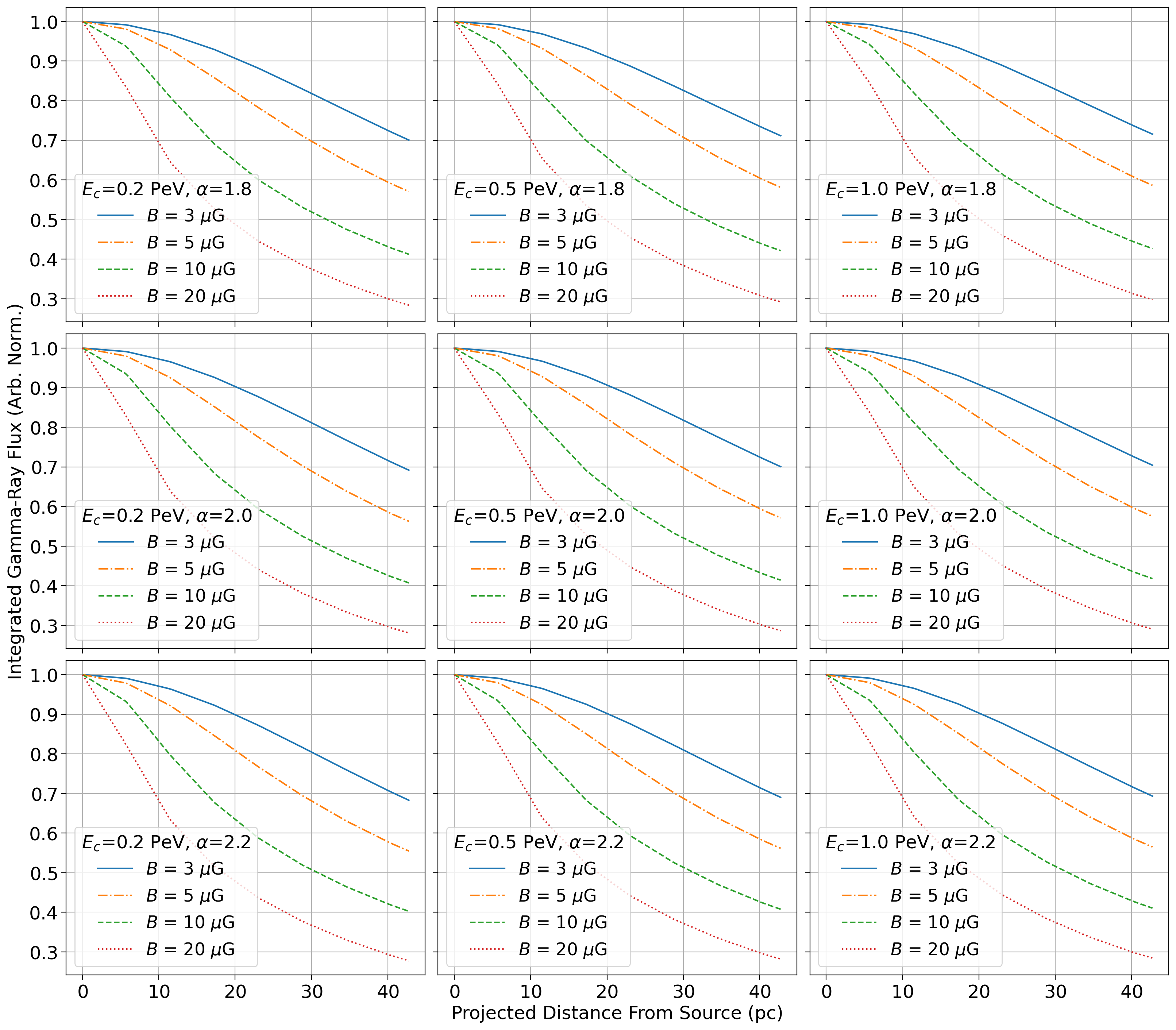}
\caption{As in Fig.~\ref{fig:radprofexample}, but adopting other values for the spectral index, $\alpha$, and exponential cutoff, $E_c$, of the injected electrons.}
\label{fig:genericradprof}
\end{figure}

\end{document}